\documentclass[aps,twocolumn,prl,showpacs,preprintnumbers,
superscriptaddress]{revtex4-1}

\usepackage{amsmath,amssymb,bm,comment}
\usepackage{graphicx}
\usepackage{epsfig}
\usepackage{epstopdf}
\usepackage{hyperref}

\begin{document}

\preprint{INT-PUB-10-011}
\title{Holographic Berezinskii-Kosterlitz-Thouless Transitions}
\author{Kristan Jensen}
\author{Andreas Karch}
\affiliation{Department of Physics, University of Washington, Seattle,
WA 98195-1560, USA}

\author{Dam T.~Son}
\affiliation{Institute for Nuclear Theory, University of Washington, Seattle,
WA 98195-1550, USA}
\author{Ethan G.~Thompson}
\affiliation{Institute for Nuclear Theory, University of Washington, Seattle,
WA 98195-1550, USA}
\date{February 2010}
\begin{abstract}
We find the first example of a quantum
Berenzinskii-Kosterlitz-Thouless (BKT) phase transition in two
spatial dimensions via holography.  This transition occurs in
the D3/D5 system at nonzero density and magnetic field.  At any
nonzero temperature, the BKT scaling is destroyed and the
transition becomes second order with mean-field exponents.
We go on to conjecture about
the generality of quantum BKT transitions in two spatial dimensions.
\end{abstract}

\maketitle

\emph{Introduction.}---%
Holography~\cite{Maldacena:1997re,Gubser:1998bc,Witten:1998qj}
has become an important tool in the
investigation of strongly-coupled systems.  Despite being restricted
to a special class of theories---those with simple gravitational
duals---the technique has found interesting applications to
the physics of the quark-gluon plasma~\cite{Son:2007vk}.
Recently, there have been
attempts to use holographic models to approach condensed matter
systems---like the Fermi gas at
unitarity~\cite{Son:2008ye,Balasubramanian:2008dm},
superfluids~\cite{Hartnoll:2008vx} and non-Fermi
liquids~\cite{Liu:2009dm,Cubrovic:2009ye,Faulkner:2009wj,Hartnoll:2009ns}---%
with various degrees of success.

Phase transitions are one central physical concept that can be studied 
holographically.
They are present in many holographic
models and frequently allow simple
geometric interpretations.  However, most holographic phase
transitions are either first-order~\cite{Hawking:1982dh,Witten:1998zw,Mateos:2006nu}
or second-order with mean-field exponents. The reason for the mean-field
behavior is the large $N$ parameter which suppresses quantum
fluctuations in the gravity theory and allows the latter to be treated
semiclassically.  On the other hand, the most interesting questions
are usually beyond the mean-field approximation~\cite{Wilson:1973jj},
or beyond the
Landau-Ginzburg-Wilson paradigm altogether~\cite{SVBSF}.
It is very important to
investigate whether the holographic method can be used to study
these non-mean-field phase transitions.

In this Letter, we show that, even within the confines of large-$N$
field theories, another type of phase transition is
possible---namely, those with the scaling behavior of the
Berezinskii-Kosterlitz-Thouless (BKT) phase
transition~\cite{Berezinskii,Kosterlitz:1973xp}.
Recall that in the BKT phase transition, the order parameter scales as
$\exp(-c/\sqrt{T_c-T})$ near the critical temperature
$T_c$~\cite{Kosterlitz:1974sm}.
In our case, interestingly, the phase transition is a
quantum phase transition, occurring at
zero temperature in 2+1 spacetime dimensions.  The explicit example
we consider is a 2+1 dimensional theory at a finite density $d$ of a
conserved charge and a magnetic field $B$.  At a particular value of
the ``filling fraction'' $\nu= d/B$, the system suffers a transition to a
broken symmetry state, and the scaling of the order parameter is the
same as in a BKT phase transition,
$\sigma\sim\exp(-c/\sqrt{\nu_c-\nu})$.
We shall call this phase
transition the ``holographic BKT phase transition,'' although it
happens in a context different from the original BKT phase transition.
The BKT scaling occurs in quantum mechanics with a $1/r^2$
potential, and
has been speculated to describe the chiral phase transition in QCD
with large number of flavors~\cite{Kaplan:2009kr}.

On the gravitational side, the transition
occurs due to the violation of the
Breitenlohner-Freedman bound~\cite{Breitenlohner:1982jf}
in the infrared region by the scalar
field dual to the order parameter\footnote{While finishing this work, we were informed of another work in progress~\cite{MIT:BKT} where the authors realize a BKT transition by a similar mechanism.}.
Although this mechanism seems similar to that of a second-order 
Landau phase transition, the field in the bulk actually
represents an infinite tower of states in the boundary quantum
field theory.  At the phase transition,
an infinite number of field theory modes
become unstable at the same time---an
extremely unnatural situation within the Landau theory.

The gravitational description of the holographic BKT transition is
given in terms of a probe brane minimizing its worldvolume action in a
fixed geometry. In Ref.~\cite{Jensen:2010vd}, three of us studied one 
such probe describing a 3+1 dimensional field theory. 
In 3+1 dimensions the competition between finite density
and the magnetic field gave rise to a second-order transition with
mean-field scaling.  (See Ref.~\cite{Jensen:2010vd,Evans:2010iy} for
technical details and a discussion of the related
literature.)  In this work, we will show that the same setup in
2+1 dimensions gives rise to BKT scaling.  The fact that the density
and the magnetic field have the same mass dimension in 2+1 dimensions will
be crucial.

\emph{The D3/D5 system.}---%
We write the AdS$_5\times S^5$ background metric as
\begin{multline}
g = (r^2 + y^2) (-dt^2 + d\vec{x}^2)\\ 
+\frac{1}{r^2 + y^2} ( dr^2 + r^2 d\Omega_2^2 + dy^2 + y^2 d\Omega_2^2),
\end{multline}
where $\vec{x}$ is a spatial three-vector and $d\Omega_2^2$ is the
metric of a unit-radius two-sphere.  In these coordinates the boundary
of the AdS$_5$ is located at $r^2+y^2\to\infty$ and the horizon
at $y = r = 0$.

We consider $N_f$ probe D5 branes in this geometry, making an Ansatz
that the probes wrap the first two-sphere, the ``radial coordinate''
$r$, time, and two spatial directions.  The branes have a profile
parameterized by $y=y(r)$.  The dual field theory is
$\mathcal{N}=4$ $SU(N)$ super-Yang-Mills theory at large 't Hooft
coupling $\lambda$ coupled to $N_f$ fundamental hypermultiplets along
a 2+1 dimensional defect.  This theory has a
$U(1)\times SU(2)_1\times SU(2)_2$ global symmetry,
where the two $SU(2)$ factors are chiral R-symmetries.
The field $y$ is dual to a condensate of the field theory which can 
spontaneously break the second SU(2) factor. We are interested in the
chiral phase transition between $y=0$ and $y\neq0$ configurations.

We turn on a density of the $U(1)$ flavor charge
and a magnetic field coupled to this charge.  On the gravity
side, this corresponds to turning on a field strength for the $U(1)$
gauge field dual to the current, $F=A'_t(r)dt\wedge dr+B dx^1\wedge
dx^2$.
The radial electric field $A'_t$ is supported by
charge behind the AdS horizon, so these branes must extend to the bottom of
AdS.

The Dirac-Born-Infeld (DBI) action density for these probe branes
takes the form
\begin{equation}\label{SyA}
S = -\mathcal{N}\!\int\! dr\, r^2 \sqrt{ 1 + y'^2 - A'^2_t }
\sqrt{ 1+\frac{B^2}{(r^2+y^2)^2}}\, .
\end{equation}
where $\mathcal{N}= \sqrt{\lambda} N_f N_c/(2 \pi^3)$.
The field $A_t$ only appears through derivatives, so
$d\equiv\delta S/\delta A'_t$ is $r$-independent.  It is the density of
the U(1) charge.  The action of $y$ at fixed density is obtained
by Legendre transforming~(\ref{SyA}) with respect to $A_t'$,
\begin{equation}\label{Snonlin}
\tilde{S} = - \int dr \sqrt{ 1 + y'^2 }
\sqrt{ d^2 + \mathcal{N}^2 r^4  + \frac{\mathcal{N}^2r^4B^2}{(r^2 + y^2)^2}}\,.
\end{equation}

The onset of the phase transition can be found by analyzing the stability
of small perturbations around $y=0$, which are described by the quadratic
part of~(\ref{Snonlin}),
\begin{equation}
  L \sim -\frac{\mathcal{N}}2 \sqrt{\rho^2+B^2+r^4}\, y^{\prime2}
       + \frac{\mathcal{N}B^2y^2}{r^2\sqrt{\rho^2+B^2+r^4}}\,,
\end{equation}
where $\rho=d/\mathcal{N}$.  We pause to note two features of this
Lagrangian.  Near the boundary at large $r$, $y/r$
fluctuates as a scalar with mass squared $m^2=-2$ in
AdS$_4$.  However, at small $r$ (the IR of the dual theory), $y/r$
behaves like a scalar in AdS$_2$ with $m^2=-2B^2/(B^2+\rho^2)$.
If $\rho/B<\sqrt{7}$, then the mass of $y/r$ in the IR region is below the
Breitenlohner-Freedman bound of stability for the effective AdS$_2$,
$m_{\rm BF}^2=-1/4$.  In this case the trivial embedding $y=0$ is 
unstable and the ground state should instead have $y\neq0$.  Thus, 
the chiral phase
transition occurs at the filling fraction
\begin{equation}
\nu_c=\frac{d}{B_c}=\mathcal{N}\frac{\rho}{B_c}
 =\frac{\sqrt{7\lambda}N_fN_c}{2\pi^3}.
\end{equation}

In the broken phase, the condensate can be found by solving the
equation for the embedding with appropriate boundary conditions.  We
are interested in the critical behavior near the phase transition.  In
this regime, the embedding can be found by matching the solutions in
two regions: a small-$r$ nonlinear core and a large-$r$ linear tail.

In the core region, we can set $B^2=\rho^2/7$, and neglect $r^4$ compared
to $\rho^2$.  The action then becomes
\begin{equation}\label{Sscaling}
  S \sim -\mathcal{N}\!\rho\!\int\! dr\,\sqrt{1+y^{\prime2}}
 \sqrt{1+\frac{r^4}{7(r^2+y^2)^2}}\,.
\end{equation}
Our brane embeddings, which extremize this action and obey the
boundary condition $y(0)=0$, form a one-parameter family of solutions
related by scaling: $y_\xi(r)= \xi f(r/\xi)$, where $y=f(r)$ is one
particular embedding.  At large $r$, any solution in the family goes
to the linear regime, where
\begin{equation}
  f(r)=\sqrt{r}(-a_0+a_1\log\, r), \,\,\, r\gg 1.
\end{equation}
Numerically solving the equation of motion by shooting, we find
a solution $f$ with $a_0=-.211$ and $a_1=.0585$.  A general embedding
in the core then has the asymptotics
\begin{equation}
  y_{\xi}(r)=a_1\sqrt{\xi r}\log \left(\frac{r}{r_0}\right), \,\,\,
  r\gg r_0=\xi e^{a_0/a_1}.
\end{equation}
It would be incorrect to use this expression for the linear tail too
far away from the core: the terms neglected when the
action~(\ref{Sscaling}) is written down become important.
But in the linear regime, the two independent solutions for $y$ can
be found without additional approximation,
\begin{equation}
\label{twosolutions}
  f_{\pm}(u) = u^{(1\pm i\alpha)/2}\, {}_2F_1 \left( \frac{1\pm i\alpha}8,
   \frac{3\pm i\alpha}8, 1\pm\frac{i\alpha}4, -u^4\right)\!,
\end{equation}
where
\begin{equation}
  \alpha = \frac{\sqrt{\rho_c^2-\rho^2}}{\sqrt{\rho^2+B^2}}\,, \qquad
  u = \frac r{(\rho^2+B^2)^{1/4}}\,.
\end{equation}
When the filling fraction is just below the phase transition, $
\alpha$ is real and small.
From the two solutions found in Eq.~(\ref{twosolutions}),
one can construct a particular linear combination
which asymptotes to $1/u$ near the boundary,
\begin{equation}
  f_{\rm n}(u) = c_+ f_+(u) + c_- f_-(u) \to \frac 1u\,, \qquad
  u\to \infty.
\end{equation}
This choice amounts to choosing zero bare mass for the dual flavor.
The coefficient of the $1/u$ term is proportional to the condensate
of the dual theory.  Denoting the condensate as $\sigma$,
the solution in the linear regime is
$y(r)=-(\rho^2+B^2)^{-1/4}\sigma f_{\rm n}(u)$.

For small $u$ and small $\alpha$, $f_{\rm n}$ can be expanded
\begin{multline}\label{fC2C1}
  f_{\rm n}(u) \sim \left[ \frac{C_1}{i\alpha}+ C_2\right] u^{(1-i\alpha)/2}
    + \left[ -\frac{C_1}{i\alpha}+ C_2\right] u^{(1+i\alpha)/2}\\
  = \sqrt u \left[ 2C_2 \cos\left(\frac\alpha 2\ln u\right)
    - \frac{2C_1}\alpha \sin\left(\frac\alpha 2\ln u\right) \right]\!,
\end{multline}
where
\begin{equation}
  C_1 = \frac{\Gamma^2(1/4)}{2^{3/4}\pi^{3/2}}\,,\quad
  C_2 = \frac{2^{1/4}\Gamma^2(5/4)(\ln256-\pi)}{\pi^{3/2}}\,.
\end{equation}
To leading order in $\alpha$, we rewrite
Eq.~(\ref{fC2C1}) as
\begin{align}
  y(r) &= \frac{2C_1\rho}{(\rho^2+B^2)^{3/8}}\sigma\frac{\sqrt r}\alpha
  \sin \left( \frac\alpha2
   \ln \frac r {r_1}\right)\!, \label{yr0}\\
   r_1 &= \left(\rho^2+B^2\right)^{1/4} \exp\left( \frac{2C_2}{C_1}\right)\!.
\end{align}

\begin{figure}
\includegraphics[scale=0.75]{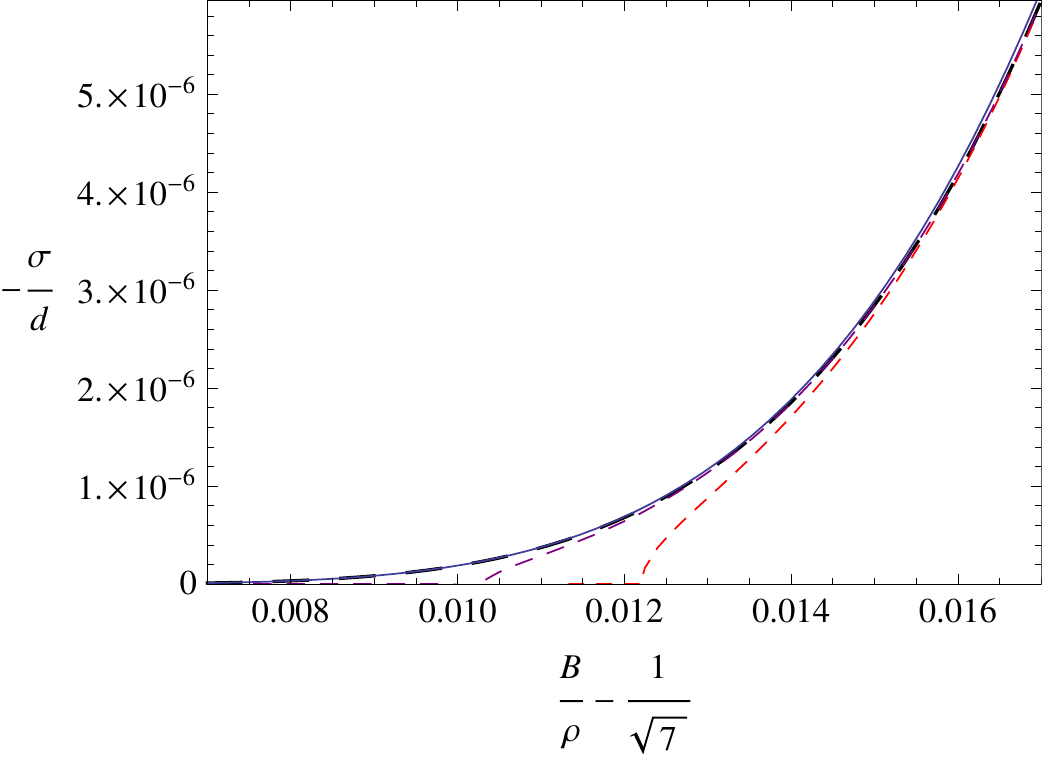}
\caption{
\label{thePlot}
A plot of the condensate as a function of magnetic field at zero and
finite temperature near the zero-temperature transition.  The dashed
black line indicates zero-temperature numerical data and
the solid blue line our prediction, Eq.~(\ref{sigmaPredict}), 
computed further to ${\cal O}(\alpha^3)$.
The color dashed curves represent numerical data
at temperatures of $T=\frac{2}{\pi}\rho^{1/2}\times 10^{-11}$ (left) and 
$T=\frac{2}{\pi}\rho^{1/2}\times 10^{-10}$ (right).
At any nonzero temperature, the
condensate scales with a mean-field exponent near the transition and
then asymptotes to the BKT scaling at large magnetic field.
}
\end{figure}

To match the core and linear solutions at $r=\xi$,
the argument of the $\sin$ in Eq.~(\ref{yr0})
has to make $\pi$ between $r_0\sim \xi$ and $r_1\sim \sqrt{B}$.  To
exponential accuracy, we then have $\xi\sim e^{-2\pi/\alpha}$.  Comparing the
prefactor of the exponent, we find
\begin{equation}
  \sigma \sim \sqrt{\xi} \sim e^{-\pi/\alpha}.
\end{equation}
Thus, the phase transition in the D3/D5 system obeys BKT scaling.
The exponent as well as the prefactors can also be found by
matching both $y$ and $y'$ at $\xi$,
\begin{equation}
\label{sigmaPredict}
  \sigma = -\mathcal{N}\frac{a_1}{C_1} \sqrt{\rho^2+B^2}
  \exp\left[ -\frac\pi\alpha + \frac{C_2}{C_1}
     - \frac{a_0}{2a_1}+O(\alpha)\right]\!.
\end{equation}
We also compare this result to numerical data in Fig.~\ref{thePlot}.

The way we matched the core and the tail of the embedding makes it
clear that there exists
an infinite set of ``Efimov extrema,'' which we name for their
resemblance to Efimov states~\cite{Efimov:1970zz}.
Indeed,
the most general match has an argument $n\pi$ in the $\sin$ of
Eq.~(\ref{yr0}) between $r\sim r_0$ and $r\sim r_1$, for $n$ a
positive integer.  These embeddings go through $n/2$ oscillations
between the bottom of the brane and the boundary.  Matching $y$ and
$y'$ at $r=\xi$ as before, we find a condensate corresponding to these
extrema of $\sigma_n\sim e^{-n\pi/\alpha}$, where $\sigma_1$ is the
condensate for the simplest embedding above.  The additional extrema
are not ground states, however: if one denotes by
$F_0$ the free energy of the trivial
embedding, the free energy of the $n$-th extremum goes like
$F_0-F_n\sim e^{-2\pi n/\alpha}$, so the $n=1$ embedding is the ground state.
The infinite tower of extrema, spaced by the same factor
$\exp (-2\pi/\alpha)$, is reminiscent of the Efimov states.

We next examine the D3/D5 system at finite temperature $T$.  To do so,
we embed the D5 probes in an AdS-Schwarzschild black hole geometry.
The embedding equations can now only be solved numerically.  However,
at any temperature the magnetic field does not contribute in the 
effective AdS$_2$
region in the IR.  We could therefore predict that the BKT scaling is
lost for $T>0$.  We numerically solved the embedding equations at a
number of different temperatures.
(The techniques employed are similar to those used in
Ref.~\cite{Jensen:2010vd}.)
We plot some of our results for the
condensate in Fig.~\ref{thePlot}.  At fixed finite temperature,
we find that there is still a chiral symmetry breaking transition but
that it is a mean-field second order one.  Far away from the
transition, we recover the exponential BKT scaling at zero
temperature.

\emph{Lifshitz systems.}---In the example that has been just considered,
the BKT scaling arises from an IR AdS$_2$ region and the existence of
a scalar with mass squared crossing the BF bound at the transition.  We
may expect the BKT scaling to be present in many other cases.  We
now demonstrate that the BKT scaling indeed controls the critical
behavior in phase transitions in a rather large class of
models.

Our models are built on the so-called Lifshitz
geometries~\cite{Kachru:2008yh}, which are invariant under anisotropic
scaling, $t\to\lambda^z t$, $\vec x\to
\lambda\vec x$.  Under holography, a geometry of this type is conjectured to be
dual to a scale-invariant boundary field theory with dynamic critical
exponent $z$.  Constructing explicit Lifshitz solutions from string
theory is quite involved~\cite{Hartnoll:2009ns}; our approach here is
phenomenological.  Nevertheless, once a few conditions (to be
specified below) are satisfied, the BKT scaling is rather generic---which
seems to indicate that it should occur in  string-theoretical
realizations of Lifshitz geometries as well.

We consider probe branes with an induced metric
\begin{equation}
\text{P}[g]=-(r^2+y^2)^zdt^2+(r^2+y^2)d\vec{x}^2+\frac{dr^2(1+y'^2)
+r^2g_{\rm int}}{r^2+y^2}
\end{equation}
and constant dilaton.  Here, $r$ is the ``radial coordinate'' of the
geometry, $z$ is the dynamical exponent of the dual theory,
$\vec{x}\in \mathbb{R}^D$ is a spatial vector, $g_{\rm int}$ is the
metric on a wrapped space of dimension $k$, and $y=y(r)$ is the
profile of the branes.  The D3/D5 system corresponds to
$z=1$, $D=2$, and $k=2$.

We now turn on a density and magnetic field in the field theory, dual
to $U(1)$ field strengths on the branes.  Assuming that
Chern-Simons terms play no role, the action of the probes at
fixed density is
\begin{equation}
\tilde{S}= -\mathcal{N}_z\!\int\! dr\, R^z\sqrt{1+y'^2}
\sqrt{\rho^2+r^{2k}R^{2(D-k)}
  \left(
  1+\frac{B^2}{R^4}
\right)
},
\end{equation}
where we have defined $R^2\equiv r^2+y^2$, $\mathcal{N}_z$ is a
prefactor, and $\rho$ is a rescaled density $d=\mathcal{N}_z\rho$.

As before, we attempt to find the location of the phase transition
by looking at small fluctuations around the trivial embedding
$y(r)=0$.
In the UV, $y/r$ is a massive scalar in AdS$_{z+3}$, corresponding to
an operator with dimension  $\Delta=z/2+\sqrt{(z+2)^2-4k}/2+1$
in field theory.
At small $r$, $y$ fluctuates in two different ways
depending on the value of $D$.  For two spatial dimensions, $y/r$
fluctuates as a $m^2=z-1-kB^2/(B^2+\rho^2)$ scalar in AdS$_{z+1}$; for
$D>2$, $y/r$ behaves like a $m^2=z-1$ scalar in AdS$_{z+1}$.  The
holographic BKT transition is then only possible for $D=2$, where a
density and magnetic field have the same dimension.  The transition
occurs at a critical density of
\begin{equation}
\frac{\rho}{B_c} = \frac{\sqrt{4k-z^2}}{z}\,,
\end{equation}
for which the bound $4k>z^2$ must also be satisfied.  This is also the
bound for which a magnetic field at zero density will break chiral
symmetry.

Using the same method as employed in the D3/D5 case, we study zero
bare mass by taking $y$ to be normalizable at large $r$ with
$y\sim1/r^{\Delta-1}$.  Holographic renormalization relates the
coefficient of this term to the condensate $\sigma$, which we find to
be
\begin{equation}
\sigma=-\mathcal{N}_z\frac{a_1}{C_1}(\rho^2+B^2)^{\frac{\Delta}{4}}
(\Delta-1)\exp \left( -\frac{\pi}{\alpha}+\frac{zC_2}{C_1}
-\frac{za_0}{2a_1}   \right)\!,
\end{equation}
where $C_1$, $C_2$ are complicated constants that depend on $\Delta$
and $z$, and $a_0$, $a_1$ are the asymptotic data of a core solution 
to which we match.  We can therefore find BKT scaling for
any dynamical exponent, provided that the magnetic field can break
chiral symmetry.

\emph{Discussion.}---%
We therefore see that holographic phase transitions in 2+1 dimensions, with the ratio
of the magnetic field and the density as a control parameter, may
generally be of the BKT type.
It would be extremely interesting to find such a
BKT transition away from the large $N$ and strong coupling limits.

The existence of an infinite number of ``Efimov vacua'' is clearly
related to the breakdown of the Landau's theory.  In Landau's theory
of phase transitions, the order parameter is the only low-energy degree
of freedom.  In our case, the role of the order parameter is played by
the whole function $y(r)$, where $r$ is usually interpreted as the energy
scale.  One needs to futher investigate the low-energy effective theory
of the D3/D5 system.

\emph{Acknowledgments.}---The authors thank Carlos Hoyos and
Steve Paik for many useful discussions.
 This work was supported in part by the U.S. Department of Energy 
under Grant Numbers DE-FG02-96ER40956 and DE-FG02-00ER41132.

\bibliography{refs}

\begin{thebibliography}{10}%
\makeatletter
\providecommand \@ifxundefined [1]{%
 \ifx #1\undefined \expandafter \@firstoftwo
 \else \expandafter \@secondoftwo
\fi
}%
\providecommand \@ifnum [1]{%
 \ifnum #1\expandafter \@firstoftwo
 \else \expandafter \@secondoftwo
\fi
}%
\providecommand \enquote [1]{``#1''}%
\providecommand \bibnamefont  [1]{#1}%
\providecommand \bibfnamefont [1]{#1}%
\providecommand \citenamefont [1]{#1}%
\providecommand\href[0]{\@sanitize\@href}%
\providecommand\@href[1]{\endgroup\@@startlink{#1}\endgroup\@@href}%
\providecommand\@@href[1]{#1\@@endlink}%
\providecommand \@sanitize [0]{\begingroup\catcode`\&12\catcode`\#12\relax}%
\@ifxundefined \pdfoutput {\@firstoftwo}{%
 \@ifnum{\z@=\pdfoutput}{\@firstoftwo}{\@secondoftwo}%
}{%
 \providecommand\@@startlink[1]{\leavevmode\special{html:<a href="#1">}}%
 \providecommand\@@endlink[0]{\special{html:</a>}}%
}{%
 \providecommand\@@startlink[1]{%
  \leavevmode
  \pdfstartlink
   attr{/Border[0 0 1 ]/H/I/C[0 1 1]}%
   user{/Subtype/Link/A<</Type/Action/S/URI/URI(#1)>>}%
  \relax
 }%
 \providecommand\@@endlink[0]{\pdfendlink}%
}%
\providecommand \url  [0]{\begingroup\@sanitize \@url }%
\providecommand \@url [1]{\endgroup\@href {#1}{\urlprefix}}%
\providecommand \urlprefix [0]{URL }%
\providecommand \Eprint[0]{\href }%
\@ifxundefined \urlstyle {%
  \providecommand \doi [1]{doi:\discretionary{}{}{}#1}%
}{%
  \providecommand \doi [0]{doi:\discretionary{}{}{}\begingroup
  \urlstyle{rm}\Url }%
}%
\providecommand \doibase [0]{http://dx.doi.org/}%
\providecommand \Doi[1]{\href{\doibase#1}}%
\providecommand \bibAnnote [3]{%
  \BibitemShut{#1}%
  \begin{quotation}\noindent
    \textsc{Key:}\ #2\\\textsc{Annotation:}\ #3%
  \end{quotation}%
}%
\providecommand \bibAnnoteFile [2]{%
  \IfFileExists{#2}{\bibAnnote {#1} {#2} {\input{#2}}}{}%
}%
\providecommand \typeout [0]{\immediate \write \m@ne }%
\providecommand \selectlanguage [0]{\@gobble}%
\providecommand \bibinfo [0]{\@secondoftwo}%
\providecommand \bibfield [0]{\@secondoftwo}%
\providecommand \translation [1]{[#1]}%
\providecommand \BibitemOpen[0]{}%
\providecommand \bibitemStop [0]{}%
\providecommand \bibitemNoStop [0]{.\EOS\space}%
\providecommand \EOS [0]{\spacefactor3000\relax}%
\providecommand \BibitemShut [1]{\csname bibitem#1\endcsname}%
\bibitem{Maldacena:1997re}%
  \BibitemOpen
  \bibfield{author}{%
  \bibinfo {author} {\bibfnamefont{J.}~\bibnamefont{Maldacena}},\ }%
  \bibfield{journal}{%
  \bibinfo {journal} {Adv. Theor. Math. Phys.}\ }%
  \textbf{\bibinfo {volume} {2}},\ \bibinfo {pages} {231} (\bibinfo {year}
  {1998})%
  \bibAnnoteFile{NoStop}{Maldacena:1997re}%
\bibitem{Gubser:1998bc}%
  \BibitemOpen
  \bibfield{author}{%
  \bibinfo {author} {\bibfnamefont{S.~S.}\ \bibnamefont{Gubser}}, \bibinfo
  {author} {\bibfnamefont{I.~R.}\ \bibnamefont{Klebanov}},\ and\ \bibinfo
  {author} {\bibfnamefont{A.~M.}\ \bibnamefont{Polyakov}},\ }%
  \bibfield{journal}{%
  \bibinfo {journal} {Phys. Lett.}\ }%
  \textbf{\bibinfo {volume} {B428}},\ \bibinfo {pages} {105} (\bibinfo {year}
  {1998})%
  \bibAnnoteFile{NoStop}{Gubser:1998bc}%
\bibitem{Witten:1998qj}%
  \BibitemOpen
  \bibfield{author}{%
  \bibinfo {author} {\bibfnamefont{E.}~\bibnamefont{Witten}},\ }%
  \bibfield{journal}{%
  \bibinfo {journal} {Adv. Theor. Math. Phys.}\ }%
  \textbf{\bibinfo {volume} {2}},\ \bibinfo {pages} {253} (\bibinfo {year}
  {1998})%
  \bibAnnoteFile{NoStop}{Witten:1998qj}%
\bibitem{Son:2007vk}%
  \BibitemOpen
  \bibfield{author}{%
  \bibinfo {author} {\bibfnamefont{D.~T.}\ \bibnamefont{Son}}\ and\ \bibinfo
  {author} {\bibfnamefont{A.~O.}\ \bibnamefont{Starinets}},\ }%
  \bibfield{journal}{%
  \bibinfo {journal} {Ann. Rev. Nucl. Part. Sci.}\ }%
  \textbf{\bibinfo {volume} {57}},\ \bibinfo {pages} {95} (\bibinfo {year}
  {2007})%
  \bibAnnoteFile{NoStop}{Son:2007vk}%
\bibitem{Son:2008ye}%
  \BibitemOpen
  \bibfield{author}{%
  \bibinfo {author} {\bibfnamefont{D.~T.}\ \bibnamefont{Son}},\ }%
  \bibfield{journal}{%
  \bibinfo {journal} {Phys. Rev.}\ }%
  \textbf{\bibinfo {volume} {D78}},\ \bibinfo {pages} {046003} (\bibinfo {year}
  {2008})%
  \bibAnnoteFile{NoStop}{Son:2008ye}%
\bibitem{Balasubramanian:2008dm}%
  \BibitemOpen
  \bibfield{author}{%
  \bibinfo {author} {\bibfnamefont{K.}~\bibnamefont{Balasubramanian}}\ and\
  \bibinfo {author} {\bibfnamefont{J.}~\bibnamefont{McGreevy}},\ }%
  \bibfield{journal}{%
  \bibinfo {journal} {Phys. Rev. Lett.}\ }%
  \textbf{\bibinfo {volume} {101}},\ \bibinfo {pages} {061601} (\bibinfo {year}
  {2008})%
  \bibAnnoteFile{NoStop}{Balasubramanian:2008dm}%
\bibitem{Hartnoll:2008vx}%
  \BibitemOpen
  \bibfield{author}{%
  \bibinfo {author} {\bibfnamefont{S.~A.}\ \bibnamefont{Hartnoll}}, \bibinfo
  {author} {\bibfnamefont{C.~P.}\ \bibnamefont{Herzog}},\ and\ \bibinfo
  {author} {\bibfnamefont{G.~T.}\ \bibnamefont{Horowitz}},\ }%
  \bibfield{journal}{%
  \bibinfo {journal} {Phys. Rev. Lett.}\ }%
  \textbf{\bibinfo {volume} {101}},\ \bibinfo {pages} {031601} (\bibinfo {year}
  {2008})%
  \bibAnnoteFile{NoStop}{Hartnoll:2008vx}%
\bibitem{Liu:2009dm}%
  \BibitemOpen
  \bibfield{author}{%
  \bibinfo {author} {\bibfnamefont{H.}~\bibnamefont{Liu}}, \bibinfo {author}
  {\bibfnamefont{J.}~\bibnamefont{McGreevy}},\ and\ \bibinfo {author}
  {\bibfnamefont{D.}~\bibnamefont{Vegh}}}%
   (\bibinfo {year} {2009}),\
  \Eprint{http://arxiv.org/abs/0903.2477}{arXiv:0903.2477 [hep-th]}%
  \bibAnnoteFile{NoStop}{Liu:2009dm}%
\bibitem{Cubrovic:2009ye}%
  \BibitemOpen
  \bibfield{author}{%
  \bibinfo {author} {\bibfnamefont{M.}~\bibnamefont{Cubrovic}}, \bibinfo
  {author} {\bibfnamefont{J.}~\bibnamefont{Zaanen}},\ and\ \bibinfo {author}
  {\bibfnamefont{K.}~\bibnamefont{Schalm}},\ }%
  \bibfield{journal}{%
  \bibinfo {journal} {Science}\ }%
  \textbf{\bibinfo {volume} {325}},\ \bibinfo {pages} {439} (\bibinfo {year}
  {2009})%
  \bibAnnoteFile{NoStop}{Cubrovic:2009ye}%
\bibitem{Faulkner:2009wj}%
  \BibitemOpen
  \bibfield{author}{%
  \bibinfo {author} {\bibfnamefont{T.}~\bibnamefont{Faulkner}}, \bibinfo
  {author} {\bibfnamefont{H.}~\bibnamefont{Liu}}, \bibinfo {author}
  {\bibfnamefont{J.}~\bibnamefont{McGreevy}},\ and\ \bibinfo {author}
  {\bibfnamefont{D.}~\bibnamefont{Vegh}}}%
   (\bibinfo {year} {2009}),\
  \Eprint{http://arxiv.org/abs/0907.2694}{arXiv:0907.2694 [hep-th]}%
  \bibAnnoteFile{NoStop}{Faulkner:2009wj}%
\bibitem{Hartnoll:2009ns}%
  \BibitemOpen
  \bibfield{author}{%
  \bibinfo {author} {\bibfnamefont{S.~A.}\ \bibnamefont{Hartnoll}}, \bibinfo
  {author} {\bibfnamefont{J.}~\bibnamefont{Polchinski}}, \bibinfo {author}
  {\bibfnamefont{E.}~\bibnamefont{Silverstein}},\ and\ \bibinfo {author}
  {\bibfnamefont{D.}~\bibnamefont{Tong}}}%
   (\bibinfo {year} {2009}),\
  \Eprint{http://arxiv.org/abs/0912.1061}{arXiv:0912.1061 [hep-th]}%
  \bibAnnoteFile{NoStop}{Hartnoll:2009ns}%
\bibitem{Hawking:1982dh}%
  \BibitemOpen
  \bibfield{author}{%
  \bibinfo {author} {\bibfnamefont{S.~W.}\ \bibnamefont{Hawking}}\ and\
  \bibinfo {author} {\bibfnamefont{D.~N.}\ \bibnamefont{Page}},\ }%
  \bibfield{journal}{%
  \Doi{10.1007/BF01208266}{\bibinfo {journal} {Commun. Math. Phys.}}\ }%
  \textbf{\bibinfo {volume} {87}},\ \bibinfo {pages} {577} (\bibinfo {year}
  {1983})%
  \bibAnnoteFile{NoStop}{Hawking:1982dh}%
\bibitem{Witten:1998zw}%
  \BibitemOpen
  \bibfield{author}{%
  \bibinfo {author} {\bibfnamefont{E.}~\bibnamefont{Witten}},\ }%
  \bibfield{journal}{%
  \bibinfo {journal} {Adv. Theor. Math. Phys.}\ }%
  \textbf{\bibinfo {volume} {2}},\ \bibinfo {pages} {505} (\bibinfo {year}
  {1998}),\ \Eprint{http://arxiv.org/abs/hep-th/9803131}{arXiv:hep-th/9803131}%
  \bibAnnoteFile{NoStop}{Witten:1998zw}%
\bibitem{Mateos:2006nu}%
  \BibitemOpen
  \bibfield{author}{%
  \bibinfo {author} {\bibfnamefont{D.}~\bibnamefont{Mateos}}, \bibinfo {author}
  {\bibfnamefont{R.~C.}\ \bibnamefont{Myers}},\ and\ \bibinfo {author}
  {\bibfnamefont{R.~M.}\ \bibnamefont{Thomson}},\ }%
  \bibfield{journal}{%
  \bibinfo {journal} {Phys. Rev. Lett.}\ }%
  \textbf{\bibinfo {volume} {97}},\ \bibinfo {pages} {091601} (\bibinfo {year}
  {2006})%
  \bibAnnoteFile{NoStop}{Mateos:2006nu}%
\bibitem{Wilson:1973jj}%
  \BibitemOpen
  \bibfield{author}{%
  \bibinfo {author} {\bibfnamefont{K.~G.}\ \bibnamefont{Wilson}}\ and\ \bibinfo
  {author} {\bibfnamefont{J.~B.}\ \bibnamefont{Kogut}},\ }%
  \bibfield{journal}{%
  \Doi{10.1016/0370-1573(74)90023-4}{\bibinfo {journal} {Phys. Rep.}}\ }%
  \textbf{\bibinfo {volume} {12}},\ \bibinfo {pages} {75} (\bibinfo {year}
  {1974})%
  \bibAnnoteFile{NoStop}{Wilson:1973jj}%
\bibitem{SVBSF}%
  \BibitemOpen
  \bibfield{author}{%
  \bibinfo {author} {\bibfnamefont{T.}~\bibnamefont{Senthil}}, \bibinfo
  {author} {\bibfnamefont{A.}~\bibnamefont{Vishwanath}}, \bibinfo {author}
  {\bibfnamefont{L.}~\bibnamefont{Balents}}, \bibinfo {author}
  {\bibfnamefont{S.}~\bibnamefont{Sachdev}},\ and\ \bibinfo {author}
  {\bibfnamefont{M.~P.~A.}\ \bibnamefont{Fisher}},\ }%
  \bibfield{journal}{%
  \bibinfo {journal} {Science}\ }%
  \textbf{\bibinfo {volume} {303}},\ \bibinfo {pages} {1490} (\bibinfo {year}
  {2004})%
  \bibAnnoteFile{NoStop}{SVBSF}%
\bibitem{Berezinskii}%
  \BibitemOpen
  \bibfield{author}{%
  \bibinfo {author} {\bibfnamefont{V.~L.}\ \bibnamefont{Berezinskii}},\ }%
  \bibfield{journal}{%
  \bibinfo {journal} {Zh. Eksp. Teor. Fiz.}\ }%
  \textbf{\bibinfo {volume} {59}},\ \bibinfo {pages} {907} (\bibinfo {year}
  {1970})%
  \bibAnnoteFile{NoStop}{Berezinskii}%
\bibitem{Kosterlitz:1973xp}%
  \BibitemOpen
  \bibfield{author}{%
  \bibinfo {author} {\bibfnamefont{J.~M.}\ \bibnamefont{Kosterlitz}}\ and\
  \bibinfo {author} {\bibfnamefont{D.~J.}\ \bibnamefont{Thouless}},\ }%
  \bibfield{journal}{%
  \bibinfo {journal} {J. Phys.}\ }%
  \textbf{\bibinfo {volume} {C6}},\ \bibinfo {pages} {1181} (\bibinfo {year}
  {1973})%
  \bibAnnoteFile{NoStop}{Kosterlitz:1973xp}%
\bibitem{Kosterlitz:1974sm}%
  \BibitemOpen
  \bibfield{author}{%
  \bibinfo {author} {\bibfnamefont{J.~M.}\ \bibnamefont{Kosterlitz}},\ }%
  \bibfield{journal}{%
  \bibinfo {journal} {J. Phys. C}\ }%
  \textbf{\bibinfo {volume} {7}},\ \bibinfo {pages} {1046} (\bibinfo {year}
  {1974})%
  \bibAnnoteFile{NoStop}{Kosterlitz:1974sm}%
\bibitem{Kaplan:2009kr}%
  \BibitemOpen
  \bibfield{author}{%
  \bibinfo {author} {\bibfnamefont{D.~B.}\ \bibnamefont{Kaplan}}, \bibinfo
  {author} {\bibfnamefont{J.-W.}\ \bibnamefont{Lee}}, \bibinfo {author}
  {\bibfnamefont{D.~T.}\ \bibnamefont{Son}},\ and\ \bibinfo {author}
  {\bibfnamefont{M.~A.}\ \bibnamefont{Stephanov}},\ }%
  \bibfield{journal}{%
  \bibinfo {journal} {Phys. Rev.}\ }%
  \textbf{\bibinfo {volume} {D80}},\ \bibinfo {pages} {125005} (\bibinfo {year}
  {2009})%
  \bibAnnoteFile{NoStop}{Kaplan:2009kr}%
\bibitem{Breitenlohner:1982jf}%
  \BibitemOpen
  \bibfield{author}{%
  \bibinfo {author} {\bibfnamefont{P.}~\bibnamefont{Breitenlohner}}\ and\
  \bibinfo {author} {\bibfnamefont{D.~Z.}\ \bibnamefont{Freedman}},\ }%
  \bibfield{journal}{%
  \Doi{10.1016/0003-4916(82)90116-6}{\bibinfo {journal} {Ann. Phys.}}\ }%
  \textbf{\bibinfo {volume} {144}},\ \bibinfo {pages} {249} (\bibinfo {year}
  {1982})%
  \bibAnnoteFile{NoStop}{Breitenlohner:1982jf}%
\bibitem{Note1}%
  \BibitemOpen
  \bibinfo {note} {While finishing this work, we were informed of another work
  in progress~\cite {MIT:BKT} where the authors realize a BKT transition by a
  similar mechanism.}%
  \bibAnnoteFile{Stop}{Note1}%
\bibitem{Jensen:2010vd}%
  \BibitemOpen
  \bibfield{author}{%
  \bibinfo {author} {\bibfnamefont{K.}~\bibnamefont{Jensen}}, \bibinfo {author}
  {\bibfnamefont{A.}~\bibnamefont{Karch}},\ and\ \bibinfo {author}
  {\bibfnamefont{E.~G.}\ \bibnamefont{Thompson}}}%
   (\bibinfo {year} {2010}),\
  \Eprint{http://arxiv.org/abs/1002.2447}{arXiv:1002.2447 [hep-th]}%
  \bibAnnoteFile{NoStop}{Jensen:2010vd}%
\bibitem{Evans:2010iy}%
  \BibitemOpen
  \bibfield{author}{%
  \bibinfo {author} {\bibfnamefont{N.}~\bibnamefont{Evans}}, \bibinfo {author}
  {\bibfnamefont{A.}~\bibnamefont{Gebauer}}, \bibinfo {author}
  {\bibfnamefont{K.-Y.}\ \bibnamefont{Kim}},\ and\ \bibinfo {author}
  {\bibfnamefont{M.}~\bibnamefont{Magou}}}%
   (\bibinfo {year} {2010}),\
  \Eprint{http://arxiv.org/abs/1002.1885}{arXiv:1002.1885 [hep-th]}%
  \bibAnnoteFile{NoStop}{Evans:2010iy}%
\bibitem{Efimov:1970zz}%
  \BibitemOpen
  \bibfield{author}{%
  \bibinfo {author} {\bibfnamefont{V.}~\bibnamefont{Efimov}},\ }%
  \bibfield{journal}{%
  \bibinfo {journal} {Phys. Lett.}\ }%
  \textbf{\bibinfo {volume} {B33}},\ \bibinfo {pages} {563} (\bibinfo {year}
  {1970})%
  \bibAnnoteFile{NoStop}{Efimov:1970zz}%
\bibitem{Kachru:2008yh}%
  \BibitemOpen
  \bibfield{author}{%
  \bibinfo {author} {\bibfnamefont{S.}~\bibnamefont{Kachru}}, \bibinfo {author}
  {\bibfnamefont{X.}~\bibnamefont{Liu}},\ and\ \bibinfo {author}
  {\bibfnamefont{M.}~\bibnamefont{Mulligan}},\ }%
  \bibfield{journal}{%
  \bibinfo {journal} {Phys. Rev.}\ }%
  \textbf{\bibinfo {volume} {D78}},\ \bibinfo {pages} {106005} (\bibinfo {year}
  {2008})%
  \bibAnnoteFile{NoStop}{Kachru:2008yh}%
\bibitem{MIT:BKT}%
  \BibitemOpen
  \bibfield{author}{%
  \bibinfo {author} {\bibfnamefont{N.}~\bibnamefont{Iqbal}}, \bibinfo {author}
  {\bibfnamefont{H.}~\bibnamefont{Liu}}, \bibinfo {author}
  {\bibfnamefont{M.}~\bibnamefont{Mezei}},\ and\ \bibinfo {author}
  {\bibfnamefont{Q.}~\bibnamefont{Si}},\ }%
  \bibinfo {journal} {{To be published}}%
  \bibAnnoteFile{NoStop}{MIT:BKT}%
\end{thebibliography}%

\end{document}